\documentclass[%
 reprint,
 superscriptaddress,
showpacs,
 amsmath,amssymb,
 aps,%onecolumn,
 twocols,
 prl
]{revtex4-1}

\usepackage{graphicx}
\usepackage{dcolumn}
\usepackage{bm}
\usepackage{color}
\usepackage{braket}
\usepackage{nicefrac}
\usepackage{booktabs}
\usepackage[colorlinks=true,linkcolor=blue,citecolor=blue,urlcolor=blue]{hyperref}

\begin{document}

\title{From classical to quantum and back: \\Hamiltonian coupling of classical and Path Integral models of atoms}

\author{Karsten Kreis}
\email[]{kreis@mpip-mainz.mpg.de}
\affiliation{Max Planck Institute for Polymer Research, Ackermannweg 10, 55128 Mainz, Germany}
\affiliation{Graduate School Materials Science in Mainz, Staudinger Weg 9, 55128 Mainz, Germany}
\author{Mark E. Tuckerman}
\email[]{mark.tuckerman@nyu.edu}
\affiliation{Department of Chemistry, New York University (NYU), New York, NY 10003, USA}
\affiliation{Courant Institute of Mathematical Sciences, NYU, New York, NY 10012, USA}
\affiliation{NYU-East China Normal University Center for Computational Chemistry at NYU Shanghai, Shanghai 200062, China}
\author{Davide Donadio}
\email[]{donadio@mpip-mainz.mpg.de}
\affiliation{Max Planck Institute for Polymer Research, Ackermannweg 10, 55128 Mainz, Germany}
\author{Kurt Kremer}
\email[]{kremer@mpip-mainz.mpg.de}
\affiliation{Max Planck Institute for Polymer Research, Ackermannweg 10, 55128 Mainz, Germany}
\author{Raffaello Potestio}
\email[]{potestio@mpip-mainz.mpg.de}
\affiliation{Max Planck Institute for Polymer Research, Ackermannweg 10, 55128 Mainz, Germany}

\date{\today}

\begin{abstract}
In computer simulations, quantum delocalization of atomic nuclei can be
modeled making use of the Path Integral (PI) formulation of quantum statistical mechanics. This approach, however, comes with a large
computational cost. By restricting the PI modeling to a small region of space,
this cost can be significantly reduced. In the present work we derive a
Hamiltonian formulation for a bottom-up, theoretically solid simulation
protocol that allows molecules to change their resolution from
quantum-mechanical to classical and {\it vice versa} on the fly, while freely
diffusing across the system. This approach renders possible simulations
of quantum systems at constant chemical potential. The validity of the proposed scheme is demonstrated by
means of simulations of low temperature parahydrogen. Potential future
applications include simulations of biomolecules, membranes, and
interfaces.
\end{abstract}


\maketitle

\section{Introduction} \label{intro}

Nuclear quantum delocalization plays a crucial role in low temperature
systems, e.g. helium or hydrogen
\cite{PhysRevB.56.14620,RevModPhys.67.279,Nozieres:441039,scharf1993,paraH2_2009,lindenauBook},
which can undergo a superfluid transition, and it affects in nontrivial ways a
large variety of systems and processes at more standard thermodynamic
conditions. It is the case, for example, for proton transfer in biomolecules and membranes
and in DNA oxidation
\cite{Smirnov2011,Jeuken2007,Haines2001,Paula1996,Lowdin:1963wt,Rein:1964ta,Perez:2010gq,Jacquemin:2014iv}, the
thermodynamics of ice \cite{Pamuk:2012jz}, the structure of water
adlayers on catalysts \cite{Li:2010fr,PhysRevLett.109.226101}, and the structure and dynamics of bulk water at room temperature \cite{Marx:1999vt,morrone_car_2008,paesaniJPCL2010,Ceriotti24092013,fritschJCTC2014}.
In order to account for these effects in computer
simulations, one can make use of Feynman's Path Integral (PI) formulation of
quantum statistical mechanics
\cite{feynman2,RevModPhys.67.279,tuckerman_review_2002,tuckermann_book}, which
enables the accurate description of nuclear delocalization by means of Monte
Carlo (MC) or Molecular Dynamics (MD) simulations
\cite{RevModPhys.67.279,tuckerman_review_2002,tuckermann_book}. This
possibility, though, comes at the expense of an increased computational
cost. A strategy to overcome this limitation is to restrict the PI description
of the atoms to a (small) region of space, where their quantum nature has to
be explicitly accounted for and to model the other atoms as classical
particles interacting {\it via} an appropriately chosen effective
potential. Molecules diffusing across the boundary separating these two
regions must change ``on the fly'' their representation from 
classical or quantum and vice versa. 
This approach, which is obviously viable only for sufficiently short De
Broglie wavelength, is beneficial especially for applications where one has,
at the same time, a small region which requires a PI description in a much larger simulation box.
Textbook examples of such systems are given by liquid-solid or liquid-liquid
interfaces \cite{gelfand1990,aguado2001,schmitz2014} and in protein
simulations \cite{hwang1996,engel2012,wang2014} in which a
chemically accurate model of the active site can be concurrently employed
with a coarser description of the rest of the molecule. The simplified
model in the classical region can also allow one to change on the fly the
number of molecules in the system \cite{debashish}, thereby implementing a
grand canonical PI approach. 
More generally, an approach in which a classical and a PI model of the system
are concurrently used in the same setup would allow a substantial
computational gain. In turn, this enables the simulation of significantly
longer length scales and sampling times compared to fully quantum PI simulations.

A first step in this direction was taken in the framework of the Adaptive Resolution Simulation (AdResS)
scheme \cite{jcp,annurev,FritschPRL} by merging quantum and classical
effective forces \cite{adolfoprl,poma_pccp,potestio}. This work demonstrated the possibility of
investigating the properties of a system of light atoms or molecules by
explicitly considering their quantum nature only locally, without disrupting
the overall thermodynamic balance between the quantum and the classical
regions. However, the AdResS scheme is intrinsically based on the
interpolation of forces, and does not admit a Hamiltonian formulation
\cite{prelu}; therefore the quantum-classical coupling was introduced {\it ad hoc} \textit{after} the quantization
of the system and the introduction of fictitious momentum coordinates. This
approach is thus incompatible with a proper PI quantization.

\begin{figure}[h!]
  \includegraphics[width=\columnwidth]{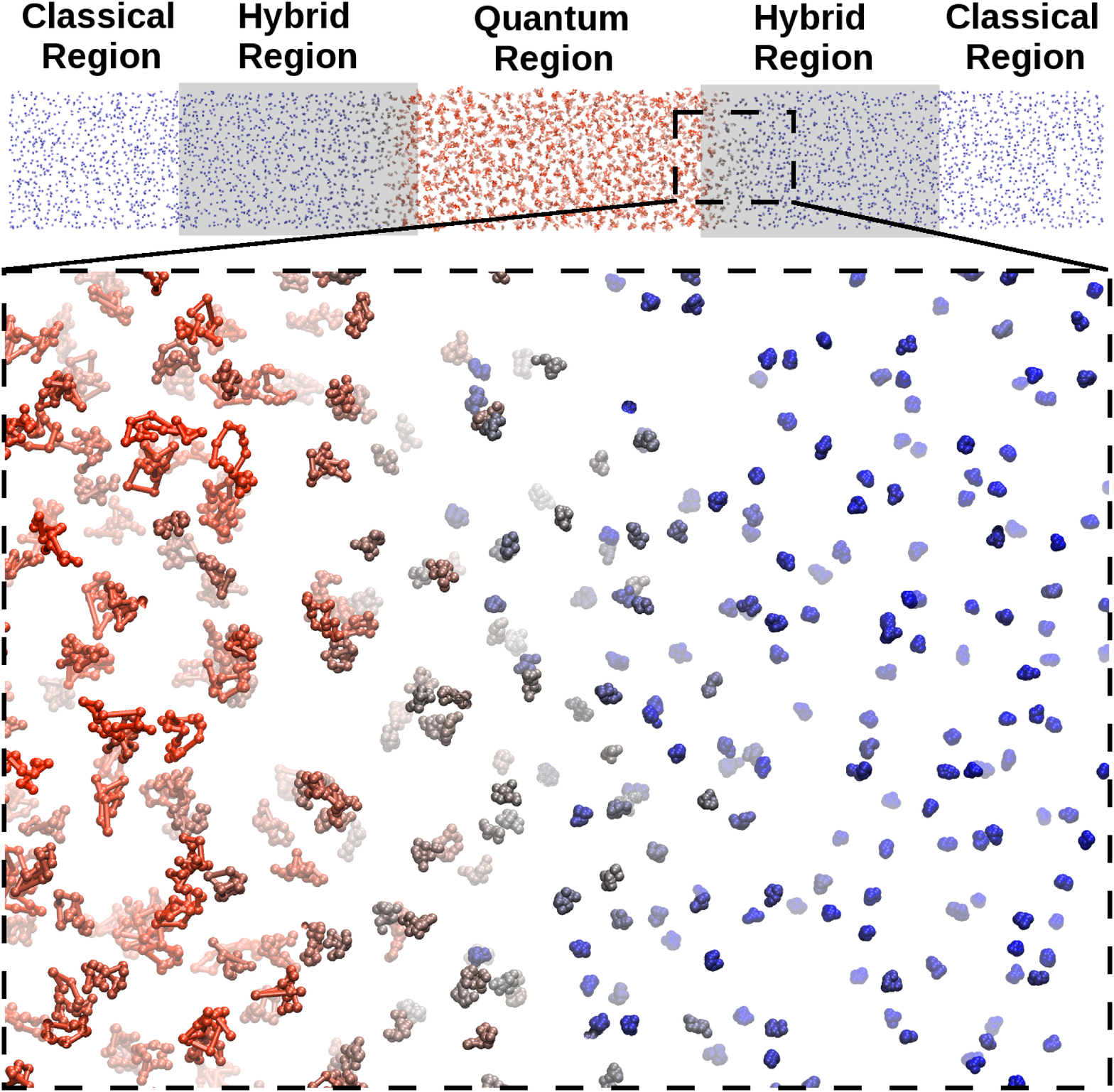}
  \caption{Illustration of the simulation setup for the quantum-classical simulations. Red (resp. blue) color corresponds to a larger (resp. smaller) radius of gyration. The smooth transition from extended to collapsed molecules demonstrates the transition from quantum mechanical to classical behavior. The particles freely move between the regions and change their behavior accordingly.\label{fig1}}
\end{figure}

In this paper, we provide a theoretically solid quantum-classical coupling
protocol, based on a global Hamiltonian. We simulate a system
of atoms or molecules that exhibit quantum behavior only in a restricted region of space and behave as purely classical particles everywhere else. Additionally, we allow molecules to freely diffuse across the simulation domain and switch the nature of their interactions according to their position in space.

\section{Adaptive Quantum-Classical Coupling} \label{adaptiveqmcl}

In order to model each subsystem 
with its appropriate interactions, we make use of the
Hamiltonian AdResS (H--AdResS) method
\cite{hadress,MC_hadress,kreisEPL2014}. This scheme was developed to perform
adaptive resolution MD/MC simulations based on a global Hamiltonian, which
makes it the appropriate framework for our work. A typical H--AdResS system is
partitioned in two regions connected {\it via} a hybrid buffer region. The
resolution of a molecule depends on the value of a representative coordinate ${\bf R}$ (usually chosen to be
the center of mass), and is parametrized by a continuous function
$\lambda({\bf R})$ smoothly switching from $0$ to $1$ in the hybrid
region. The total potential energy of each molecule is obtained by 
interpolating between the two resolutions. The H-AdResS Hamiltonian $H$ of a system of point-like particles reads
\begin{eqnarray}\label{hadress_H}
H = \mathcal K + \sum_{\alpha=1}^N \left[{\lambda_\alpha} {V^{1}_\alpha} + {(1 - \lambda_\alpha)} {V^{0}_\alpha}  - \Delta H(\lambda_\alpha) \right]
\end{eqnarray}
where $\mathcal K$ is the kinetic energy,
$\alpha$ indexes the $N$ particles, 
and $\lambda_\alpha = \lambda(\textbf{R}_\alpha)$. The single-particle
potentials $V^{Res}_\alpha$ (with $Res = 0, 1$) are the sums of all
intermolecular potentials acting on particle $\alpha$, properly normalized so
that double counting is avoided \cite{hadress,MC_hadress}. In the following we
make no assumption about the specific form of these interactions.
The term $\Delta H$, referred to as the Free Energy
Compensation (FEC) \cite{hadress,MC_hadress}, is an external field acting only
in the hybrid region to neutralize the density imbalance that naturally occurs when different models of the same system are coupled
together. Its calculation is described 
in the {\it Validation} section.

The employment of the H--AdResS Hamiltonian in the PI formalism is
straightforward. Specifically, the ring polymer potential energy obtained from
the PI quantization of the Hamiltonian in Eq. (\ref{hadress_H}),
assuming Boltzmann statistics, is given by
\begin{equation}\label{vp:1}
\begin{split}
V_P &= \sum_{\alpha = 1}^N\ \sum_{k = 1}^P \left\{ \frac{m_{\alpha} \omega_P^2}{2} |{\bf r}_{\alpha,k} - {\bf r}_{\alpha,k+1}|^2 \right.\\
&+\left. \frac{1}{P} \left [\lambda_{\alpha,k} V^{1}_{\alpha,k} + (1 - \lambda_{\alpha,k}) V^{0}_{\alpha,k}   - \Delta H(\lambda_{\alpha,k})\right ] \right\}
\end{split}
\end{equation}
for $N$ interacting particles in 3 dimensions, where $\omega_P \equiv \sqrt{P}
/ \beta \hbar$, $\beta = 1/k_B T$, $T$ is the temperature, $k_B$ is
Boltzmann's constant, and $\hbar$ is Planck's constant. The index $k$ labels
the $P$ ``copies'' of the original system, which, after quantization consists of $N$ 
ring-polymers, each containing $P$ beads, $\lambda_{\alpha,k} = \lambda({\bf r}_{\alpha,k})$, and $V^{Res}_{\alpha,k}$ is the total interaction of type $Res = 0,1$ on replica $k$ of particle $\alpha$.

Eq. (\ref{vp:1}) describes a system of quantum particles,
represented by ring polymers whose interactions change in space. Nevertheless,
their quantum behavior, dictated by the strength of the springs connecting the
beads of each ring, is the same everywhere. At this stage we need a strategy
to switch between the quantum and classical
descriptions of the particles. This can be achieved by
modifying the mass of the atoms, as larger masses
correspond to stronger springs of elastic constant $m \omega_P^2$; a
large mass causes the ring polymers collapse, and the particles
approach their classical limit. We thus define
\begin{equation}\label{eq:mass}
m \rightarrow \mu(\lambda) = \lambda m + (1 - \lambda) M
\end{equation}
where $\mu(\lambda)$ smoothly switches from a mass $\mu(0) = M$ to a mass
$\mu(1) = m \ll M$. For $\mu = m$, which is set to be the real, physical mass,
the particles are light and the quantum zero-point motion
becomes important. In contrast, 
the mass $M$ should be large enough to give
the particles a classical character.

We now proceed with the quantization of a system of particles with
position-dependent masses. As a starting point we consider the
Hamiltonian operator for a free particle of mass $\mu(x)$ in one dimension
(the procedure generalizes trivially to many-particle systems in three
dimensions). The Hamiltonian must be represented as a 
{\it Hermitian} operator, which we can obtain by writing it in the 
following form:
\begin{equation}\label{eq:H1}
\hat{\mathcal{H}} = \frac{1}{2}\hat{p} \mu^{-1}(\hat{x})\hat{p}
\end{equation}
where $\mu^{-1}(\hat{x})$ is the inverse mass operator, and $\hat{p}$ is the momentum operator. Using this Hamiltonian, we seek to formulate the partition function $Q=\text{Tr}[\exp\{-\beta\hat{\mathcal{H}}\}]$ as a path integral. Introducing the usual set of $P$ resolutions of the identity operator, we can write the trace as
\begin{equation}\label{eq:H2}
\begin{split}
Q=&\lim\limits_{P\to\infty}\int dx_1\cdots dx_P \\
&\prod_{k=1}^P\bra{x_k}\exp\left( -\frac{\beta}{2P}\hat{p} \mu^{-1}(\hat{x})\hat{p} \right)\ket{x_{k+1}}\bigg|_{x_{P+1}=x_1}
\end{split}
\end{equation}

Note that in the free particle case, it is not necessary to make use of the
limit $P\rightarrow\infty$, required when applying Trotter's theorem. However,
the latter is generally necessary in presence of a potential $V(\hat{x})$, hence we introduce this limit at this stage without any loss of generality.

To derive the matrix elements in Eq. (\ref{eq:H2}), we introduce the momentum identity resolution:
\begin{equation}\label{eq:H3}
\begin{split}
&\bra{x_k}\exp\left( -\frac{\beta}{2P}\hat{p} \mu^{-1}(\hat{x})\hat{p} \right)\ket{x_{k+1}}\\
&=\int_{-\infty}^{\infty}dp\braket{x_k|p}\bra{p} \exp\left( -\frac{\beta}{2P}\hat{p} \mu^{-1}(\hat{x})\hat{p} \right) \ket{x_{k+1}}
\end{split}
\end{equation}
Given that the limit $P\rightarrow\infty$ is ultimately taken, we can work with an infinitesimal version of the exponential operators by expanding the exponential to first order. Thus, we obtain
\begin{equation}\label{eq:H4}
\begin{split}
&\bra{p} \exp\left( -\frac{\beta}{2P}\hat{p} \mu^{-1}(\hat{x})\hat{p} \right) \ket{x}\\
&\approx \bra{p} \left( 1 -\frac{\beta}{2P}\hat{p} \mu^{-1}(\hat{x})\hat{p} \right) \ket{x}
\end{split}
\end{equation}
Now, we introduce the commutator $[\mu^{-1}(\hat{x}),\hat{p}]$ and write
\begin{equation}\label{eq:H5}
\begin{split}
\mu^{-1}(\hat{x})\hat{p} & = \hat{p}\mu^{-1}(\hat{x}) + [\mu^{-1}(\hat{x}),\hat{p}] \\
& = \hat{p}\mu^{-1}(\hat{x}) +i\hbar\frac{d\mu^{-1}}{d\hat{x}}
\end{split}
\end{equation}
Substituting Eq. (\ref{eq:H5}) into Eq. (\ref{eq:H4}) yields
\begin{equation}\label{eq:H6}
\begin{split}
&\bra{p} \left(1 -\frac{\beta}{2P}\hat{p}^2 \mu^{-1}(\hat{x})- \frac{i\hbar\beta}{2P}\hat{p}\frac{d\mu^{-1}}{d\hat{x}} \right) \ket{x}\\
& = \braket{p|x}\left( 1 -\frac{\beta p^2}{2P}\mu^{-1}(x)- \frac{i\hbar\beta}{2P}p\frac{d\mu^{-1}}{dx} \right) \\
& \approx \braket{p|x} \exp\left[ -\frac{\beta}{2P}\left( p^2\mu^{-1}(x) + i\hbar p\frac{d\mu^{-1}}{dx}\right) \right]
\end{split}
\end{equation}
where the operators are now replaced by the corresponding eigenvalues. Substituting Eq. (\ref{eq:H6}) into Eq. (\ref{eq:H3}) gives
\begin{equation}\label{eq:H7}
\begin{split}
&\bra{x_k}\exp\left( -\frac{\beta}{2P}\hat{p} \mu^{-1}(\hat{x})\hat{p} \right)\ket{x_{k+1}}\\
&=\int_{-\infty}^{\infty}dp\braket{x_k|p}\braket{p|x_{k+1}}\times \\
& \quad\qquad \times\exp\left[ -\frac{\beta}{2P}\left( p^2\mu^{-1}(x_{k+1}) + i\hbar p\frac{d\mu^{-1}}{dx}\bigg|_{x_{k+1}} \right) \right] \\
&=\left( \frac{\mu(x_{k+1})P}{2\pi\beta\hbar^2} \right)^{\frac{1}{2}} \exp \Bigg\{ -\frac{\beta\mu(x_{k+1})P}{2(\beta\hbar)^2} \bigg[  (x_k-x_{k+1}) + \\
& \quad\qquad -\frac{\beta\hbar^2}{2P}\frac{d\mu^{-1}}{dx}\bigg|_{x_{k+1}} \bigg]^2 \Bigg\}
\end{split}
\end{equation}
where the last equality has been obtained by introducing the matrix elements
$\braket{x|p} = \exp(i p x /\hbar)/\sqrt{2 \pi \hbar}$ and performing
the momentum integration by
completing the square. 

From Eq. (\ref{eq:H7}), we see that the inverse mass derivative term
can be neglected if the following condition holds:
\begin{equation}\label{eq:H8}
\left| \left( \frac{d\mu^{-1}}{dx} \right)_{x_{k+1}} \right| \ll \frac{2P\Delta x_{k,k+1}}{\beta\hbar^2}
\end{equation}
where we defined $\Delta x_{k,k+1}=|x_k-x_{k+1}|$. Since,
\begin{equation}\label{eq:H9}
\left| \left( \frac{d\mu^{-1}}{dx} \right)_{x_{k+1}} \right| = \left| \frac{1}{\mu^2(x_{k+1})} \left( \frac{d\mu}{dx} \right)_{x_{k+1}} \right|
\end{equation}
the condition becomes
\begin{equation}\label{eq:H10}
\left| \left( \frac{d\mu}{dx} \right)_{x_{k+1}} \right| \ll \frac{2\Delta x_{k,k+1}}{\Lambda^2_\mu(x_{k+1})}\mu(x_{k+1})
\end{equation}
using the definition of position-dependent De Broglie wavelength $\Lambda_\mu(x) \equiv \sqrt{\beta\hbar^2/(P\mu(x))}$. Since $\langle\Delta x\rangle\equiv\sqrt{\langle\frac{1}{P}\sum_{l=1}^P\Delta x^2_{l,l+1}\rangle}=\Lambda_\mu\sqrt{(P-1)/P}\approx\Lambda_\mu$ for a free ring of constant mass $\mu$ and typical values of P, we can approximate $\Delta x_{k,k+1}\approx \Lambda_\mu(x_{k+1})$ and write
\begin{equation}\label{eq:H11}
\left| \frac{d\mu(x)}{dx} \right| \ll \frac{2\mu(x)}{\Lambda_\mu(x)}
\end{equation}
for an arbitrary position $x$.
The inequality in Eq. (\ref{eq:H11}) must be satisfied
everywhere in the system. In the classical and quantum regions this is
trivially the case, as the resolution function $\lambda(x)$ is flat
there. This condition means that the interpolation within the hybrid region
needs to be sufficiently smooth in order to neglect the term containing the
mass gradient in Eq. (\ref{eq:H7}). This can always be achieved by utilizing a
sufficiently large coupling region. In that sense, the criterion can be
interpreted as a lower bound on the width of the hybrid region. Furthermore,
it also holds in the presence of typical potentials, since these typically do
not dramatically change the radius of
gyration and the intrabead distances of the polymer rings compared to free
rings. Additionally, although the derivation was carried out for one dimensional systems, the derivation generalizes trivially to higher dimensions. This criterion is also correct in three spatial dimensions, as the mass change only happens along one of these 
dimensions, and therefore, only the bead-bead distances projected onto this direction matter.

Concluding, if the inequality in Eq. (\ref{eq:H11}) is fulfilled, then introducing the H--AdResS potential energy, we obtain the following partition function for $N$ interacting Boltzmann particles in three dimensions:
\begin{equation}\label{eq:H12}
\begin{split}
Q&=\lim\limits_{P\to\infty}\left[ \prod_{k=1}^P\prod_{\alpha=1}^N \left( \frac{mP}{2\pi\beta\hbar^2} \right)^{\frac{3}{2}} \right]\times \\
&\quad\qquad \times\int\prod_{k=1}^P\prod_{\alpha=1}^N d\boldsymbol{r}_{\alpha,k}\exp\left\{ -\beta V_P^\mu \right\}
\end{split}
\end{equation}
with 
\begin{equation}\label{vp:2}
\begin{split}
V_P^\mu &= \sum_{k = 1}^P\ \sum_{\alpha = 1}^N \left\{ \frac{\mu_{\alpha,k}\ \omega_P^2}{2} |{\bf r}_{\alpha,k} - {\bf r}_{\alpha,k+1}|^2  - \frac{3}{2\beta} \log{\frac{\mu_{\alpha,k}}{m}}\right.\\
&+\left. \frac{1}{P} \left [\lambda_{\alpha,k} V^{1}_{\alpha,k} + (1 - \lambda_{\alpha,k}) V^{0}_{\alpha,k}   - \Delta H(\lambda_{\alpha,k}) \right] \right\}
\end{split}
\end{equation}
and $\mu_{\alpha,l} = \mu({\bf r}_{\alpha,l})$. In Eq. (\ref{vp:2}) the position-dependent normalization prefactor has been explicitly introduced in the potential $V_P^\mu$ as a logarithmic function of the bead masses, so that it can be treated as a conventional energy term and fully removed from the Hamiltonian by means of the FEC function $\Delta H$ in Eq. (\ref{hadress_H}), in a manner similar to that done in Ref. \cite{potestioEPJB2014}. The light mass $m$ has been used as the reference mass scale. A different choice would not affect the final result of the calculations. Using the mass $m$ as a reference, however, the normalization prefactor corresponds to the one known for 
PIs with constant mass $m$ \cite{feynman2,tuckermann_book}. The ring polymers described by the energy function $V_P^\mu$ (Eq. (\ref{vp:2})) are expanded in the region where the mass is small and collapse to nearly 
classical point-like particles in the large-mass region.

\section{Validation} \label{validation}

To validate the proposed quantum-to-classical coupling scheme, adaptive Path Integral MC simulations of liquid parahydrogen at $20\,\text{K}$ with $P=16$ are performed. Other test cases might be considered, e.g. water at room temperature, but, in spite of the important role played by nuclear quantum effects in this example \cite{Marx:1999vt,morrone_car_2008,paesaniJPCL2010,Ceriotti24092013,fritschJCTC2014}, the hydrogen atoms feature a relatively small delocalization. Ultracold hydrogen, on the other hand, exhibits a more pronounced quantum mechanical character \cite{scharf1993,paraH2_2009,lindenauBook}. We hence study
it as an extreme case, well-suited to test the proposed algorithm.

\subsection{System setup} \label{setup}

\begin{figure}[!ht]
  \includegraphics[width=\columnwidth]{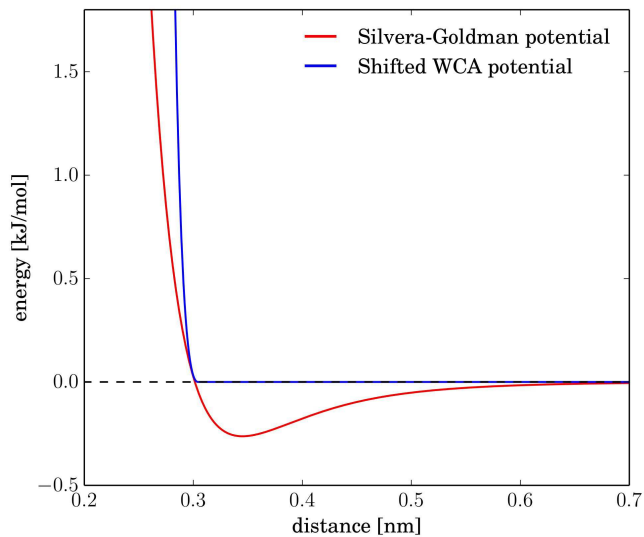}
  \caption{Non-bonded intermolecular interaction potentials used in the adaptive quantum-classical simulations. The red curve is the Silvera Goldman potential, which is employed in the low-mass quantum region. The blue curve shows the shifted WCA potential, which is used in the high-mass classical region.\label{fig1pots}}
\end{figure}

We consider a system composed of 4964 hydrogen molecules in a slab of dimensions $24.000\,\text{nm}\times 3.123\,\text{nm}\times 3.123\,\text{nm}$ (molecular density $28.4\,\text{cm}^3/\text{mol}$) with periodic boundary conditions in all directions. The width of the low-mass quantum region is set to $d_{\text{QM}}=6.0\,\text{nm}$ and the thickness of each hybrid transition region is $d_{\text{HY}}=5.0\,\text{nm}$. In order to assign to a bead its position-dependent resolution $\lambda$, its distance from the boundary between the quantum and the hybrid region is computed, i.e. $|x_{\alpha,k}|-d_{\text{QM}}/2$, where $x_{\alpha,k}$ denotes the $X$ coordinate of the bead in a coordinate system with its origin at the center of the simulation box. This quantity is then employed in the resolution function $\lambda(x)$, which is given by
\begin{equation}\label{ResFct}
\lambda(x) =
\left\{
	\begin{array}{ll}
		1 & \mbox{: } x \leq 0 \\
		\text{cos}^2\left(\frac{\pi}{2}\,\frac{x}{d_{\text{HY}}}\right) & \mbox{: } 0 < x < d_{\text{HY}} \\
		0 & \mbox{: } x \geq d_{\text{HY}}
	\end{array}
\right.
\end{equation} 
The mass $m$ is set to the molecular hydrogen mass $m_{\text{H}_2} = 2.001\,au$. In the classical region the increased mass is chosen as $M=100\,m_{\text{H}_2}$. In the quantum (QM) region we employ the Silvera-Goldman potential \cite{silvera1978,silvera1980} with a cutoff at $0.9\,\text{nm}$ for the intermolecular interaction potential $V^{1}$, while in the classical (CL) region we make use of a shifted, purely repulsive Weeks Chandler Andersen (WCA) potential \cite{wca}:
\begin{equation}
V^{0} =
\left\{
	\begin{array}{ll}
		4\epsilon\Bigl[\left(\frac{\sigma}{r-r_0}\right)^{12}-\left(\frac{\sigma}{r-r_0}\right)^6+\frac{1}{4}\Bigr]  & \mbox{if } r \leq R_{\text{c}} \\
		0 & \mbox{if } r > R_{\text{c}}
	\end{array}
\right.
\end{equation}
where $r=|\boldsymbol{r}_{\alpha,k}-\boldsymbol{r}_{\beta,k}|$, ($\alpha\neq\beta$) denotes the distance between beads of the same imaginary time slice $k$ in different molecules $\alpha$ and $\beta$. Furthermore, we choose $\epsilon=1.0\,\text{kJ/mol}$, $\sigma =0.14\,\text{nm}$, and $r_0=0.15\,\text{nm}$. The cutoff is given by $R_{\text{c}}=2^{\nicefrac{1}{6}}\sigma+r_0$. The two potentials are graphically presented in Fig. \ref{fig1pots}. This potential is not to be interpreted as a classical model for low temperature parahydrogen; rather, it was parametrized only to approximately reproduce the hard-core radius of the reference quantum particles. Other choices, suitable to other simulation setups, are clearly possible. We purposely avoid fitting the classical potential to the structure of the reference to demonstrate the generality of the protocol.

The chosen set of parameters also satisfies Eq. (\ref{eq:H11}). Finally, we stress that in the CL region the WCA interaction between ring polymers is computed only using the center of mass of the ring, thus gaining an effective reduction of the computational cost. This simplification is allowed by the essentially point-like structure of the rings in the CL regions, as can be seen from the radius of gyration profile (Fig. \ref{fig2}). The number of computations per pair of molecule is reduced from $P=16$ to one.

To modulate the thermodynamic imbalance between the classical high-mass and the low-mass quantum subsystems a Free Energy Compensation (FEC) is applied \cite{hadress,MC_hadress}. To compute the $\Delta H_{\text{KTI}}(\lambda)$ compensation a Kirkwood Thermodynamic Integration \cite{kirkwood1935} of a smaller system of 360 molecules in a box with dimensions $2.570\,\text{nm}\times 2.570\,\text{nm}\times 2.570\,\text{nm}$ is performed.

In order to remove also the remaining fluctuations in the obtained density profile after applying the Kirkwood-based FEC, an iterative approach similar to the one presented in \cite{FritschPRL} is employed. The normalized density profile $\tilde{\rho}_{\text{HY}}(x)$ in the hybrid region is transformed into a function of the resolution, $\rho_{\text{HY}}(\lambda)$, and then converted into a correction energy of the form
\begin{equation}
\Delta \tilde{H}(\lambda) = -\frac{1}{\beta}\,\text{ln}\{\rho_{\text{HY}}(\lambda)\}
\end{equation}
The latter quantity is then applied as part of the FEC $\Delta H(\lambda)$ 
in addition to 
the term $\Delta H_{\text{KTI}}(\lambda)$ obtained from Kirkwood thermodynamic integration. This is done in an iterative fashion, until a sufficiently flat density profile is obtained. The protocol for the FEC then reads
\begin{equation}
\Delta H^{i+1}(\lambda) = \Delta H^{i}(\lambda) -\frac{1}{\beta}\,\text{ln}\{\rho_{\text{HY}}^{i}(\lambda)\}
\end{equation}
with $\Delta H^{0}(\lambda)=\Delta H_{\text{KTI}}(\lambda)$ and $\rho_{\text{HY}}^{0}(\lambda)=\rho_{\text{HY}}^{\text{KTI}}(\lambda)$, where $\rho_{\text{HY}}^{\text{KTI}}(\lambda)$ corresponds to the initial hybrid region density profile obtained from simulations in which only the Kirkwood-based FEC term is applied. The protocol converges by construction when a flat density profile is achieved.

\subsection{Monte Carlo sampling} \label{montecarlo}

To sample the system's phase space we employ a standard Metropolis Monte Carlo algorithm \cite{tuckermann_book}. For the Kirkwood TI of the small system we run 16 simulations with $10^5$ sweeps each. The $\lambda$ parameter increases linearly every sweep by $10^{-5}$. The results are averaged after the simulations. Employing the Kirkwood TI FEC term thus obtained we then run 5 iterations of simulations with applying the protocol set out above to refine the density profile. Each iteration consists of 32 parallel simulations with each of these running $5\cdot10^3$ equilibration sweeps and another $5\cdot10^3$ sweeps during which the density profile is measured. Also here, after each iteration the results are averaged. Having reached a sufficiently smooth density profile, we then utilize the FEC from the Kirkwood TI and the iterative protocol to perform the main production simulations. For these we perform 32 simulations in parallel, each running $4\cdot10^3$ sweeps. Afterwards, the results (i.e. the RDF's, the density profiles as well as the radius of gyration profiles) are once again averaged over all simulations.

Each sweep is constituted by $N$ attempted Monte Carlo moves on randomly chosen molecules, with $N$ being the total number of molecules ($N=4964$ in the production run simulations). Three different kinds of moves are randomly performed:

\textbf{Whole molecule displacements}: The chosen molecule is displaced as a whole by moving its center of mass. The direction is chosen randomly from a uniform spherical distribution and the distance is drawn from a Gaussian distribution with zero mean and width $\sigma_{\text{CoM}}$.

\textbf{Molecule rotations}: The chosen molecule is rotated as a whole around a randomly oriented axis passing through its center of mass. The angle is chosen randomly from a Gaussian distribution with zero mean and width $\sigma_{\text{rot}}$.

\textbf{Individual Trotter-bead moves}: An individual bead of the molecule is randomly chosen and displaced. The direction is chosen randomly from a uniform spherical distribution and the distance is drawn from a Gaussian distribution with zero mean and width $\sigma_{\text{bead}}$. 

The different values for the $\sigma_i$'s of all simulations are presented in Tab. \ref{fig_supp_1}. In the adaptive resolution simulation, they are chosen such that they result in adequate acceptance ratios for the moves both in the classical high-mass as well as in the quantum low-mass region. When picking a molecule for a Monte Carlo move the probabilities for performing whole molecule displacements or molecule rotations are $\nicefrac{1}{13}$ each while the probability for Trotter-bead moves was $\nicefrac{11}{13}$. This choice leads to a convenient balance between whole molecule motions and Trotter-bead fluctuations.

\begin{table}[!ht]\renewcommand{\arraystretch}{1.3}\addtolength{\tabcolsep}{-1pt}\renewcommand{\tabcolsep}{0.3cm}
\begin{center}
\begin{tabular}{l c c c}
\toprule[0.03cm]
Simulation        & $\sigma_{\text{CoM}}$ & $\sigma_{\text{rot}}$ & $\sigma_{\text{bead}}$  \\ 
\midrule[0.03cm]
Kirkwood TI & $0.1\,\text{nm}$ & $0.5\,\text{rad}$ & $0.03\,\text{nm}$ \\
Adaptive Simulation & $0.1\,\text{nm}$ & $0.5\,\text{rad}$ & $0.03\,\text{nm}$  \\
QM Reference & $0.1\,\text{nm}$ & $0.5\,\text{rad}$ & $0.07\,\text{nm}$  \\
Classical Reference & $0.1\,\text{nm}$ & - & -  \\
\bottomrule[0.03cm]
\end{tabular}
\caption{Widths of the Gaussian distributions employed to draw the random displacements and rotations from for the Kirkwood Thermodynamic Integration, for the reference simulations as well as for the adaptive quantum-classical simulations.}
\label{fig_supp_1}
\end{center}
\end{table}

\subsection{Reference simulations} \label{reference}

To be able to evaluate the results of the adaptive quantum-classical simulations, we perform full-quantum as well as full-classical reference simulations of liquid parahydrogen for comparison. The systems are composed of 828 molecules in a box with dimensions $4.003\,\text{nm}\times 3.123\,\text{nm}\times 3.123\,\text{nm}$. These parameters result in the same density as in the adaptive simulations. Likewise, the temperature is set to $T=20\,\text{K}$ and the Silvera-Goldman potential is employed. For the full-quantum simulations we choose $P=16$ as in the adaptive simulations while the classical simulations are performed with $P=1$. In both cases, 16 simulations are run in parallel, each one for $2\cdot10^4$ sweeps. Afterwards the results are averaged. The values used for the $\sigma_i$'s in the reference simulations are presented in Tab. \ref{fig_supp_1}. 
In the classical simulations, all moves are as described above with the obvious exception of bead and rotating moves, which do not exist for 
classical particles.

\subsection{Results} \label{results}

\begin{figure}[h]
  \includegraphics[width=\columnwidth]{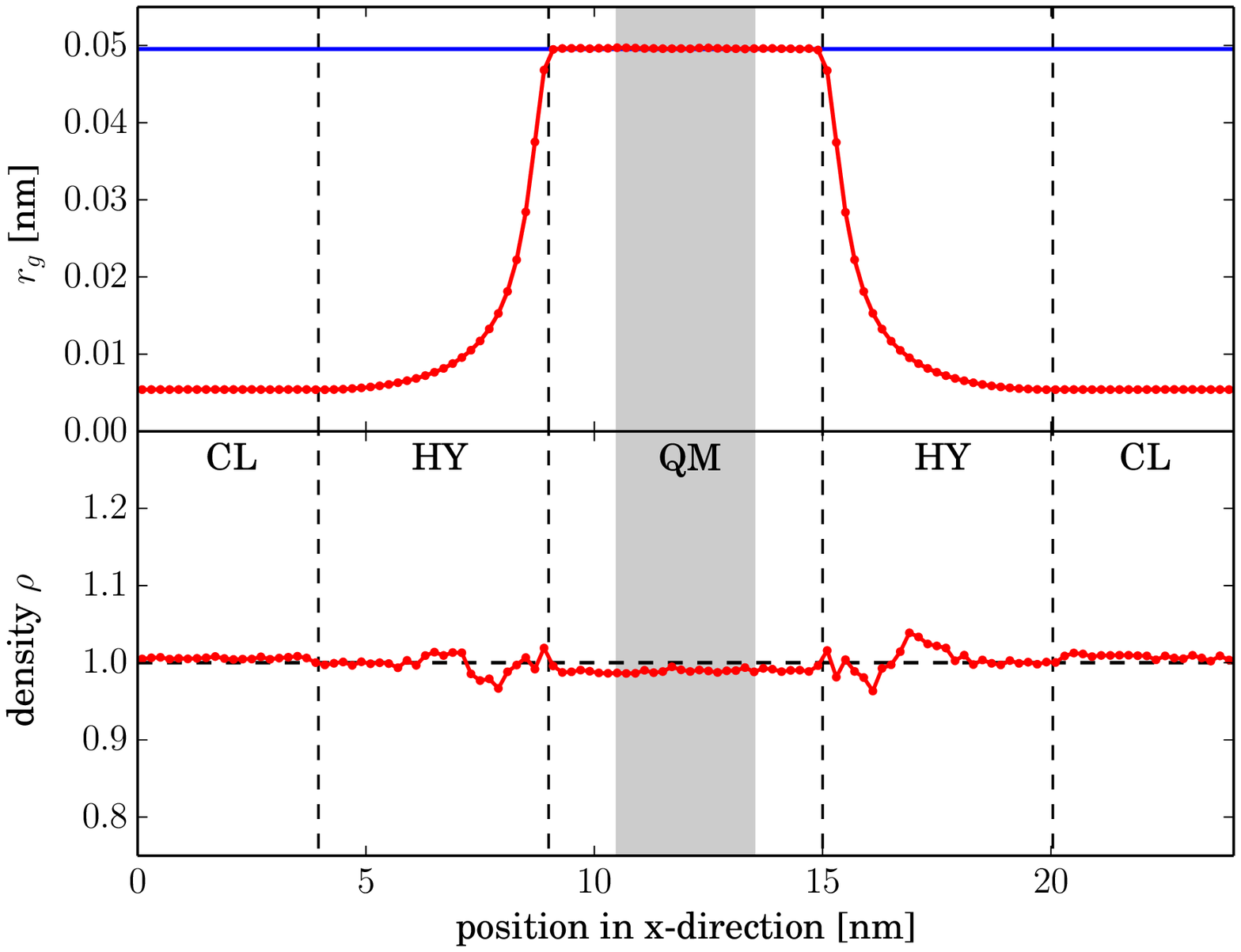}\\
	\includegraphics[width=\columnwidth]{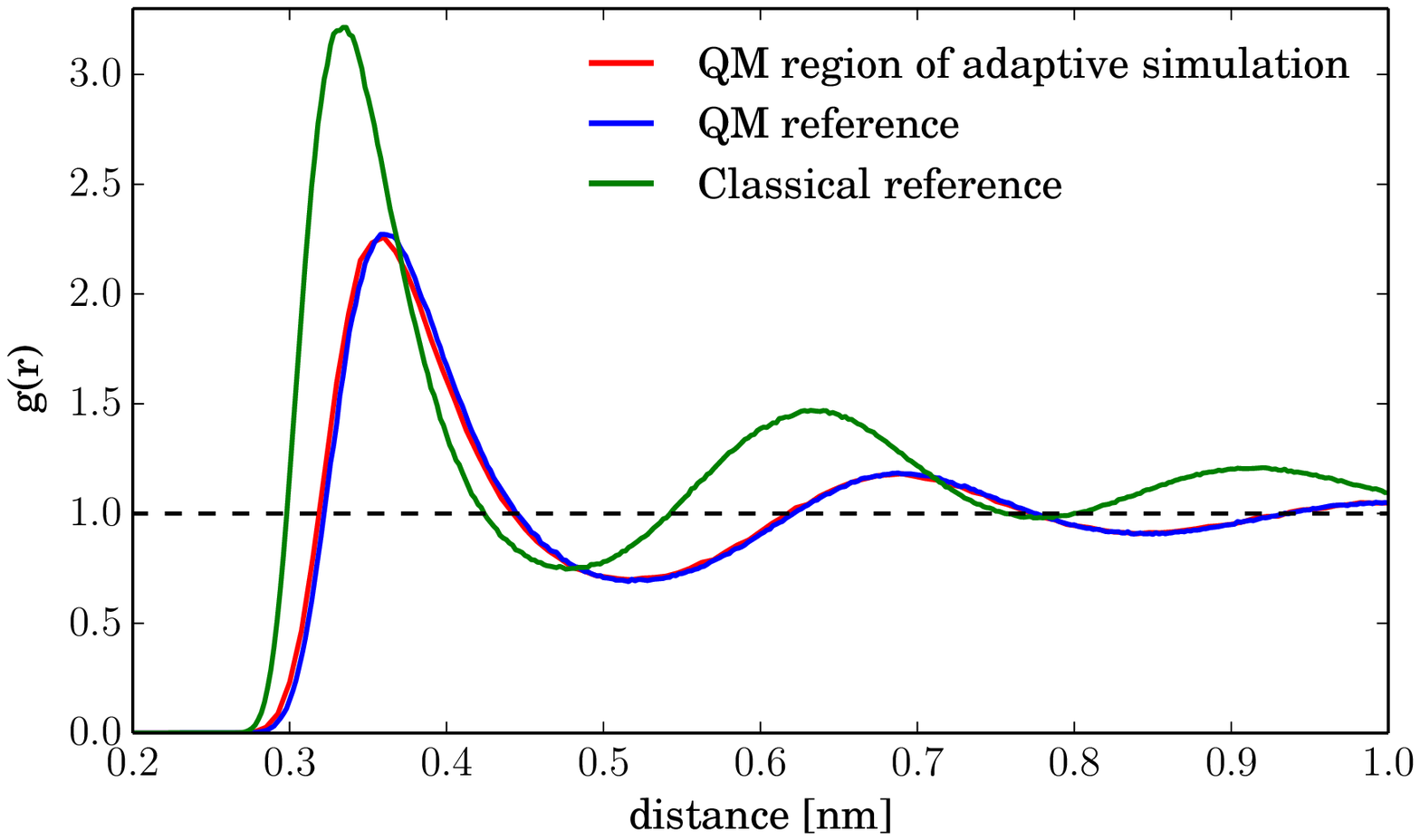}
  \caption{Top: profiles of the radius of gyration $r_g$ and the normalized density $\rho$. The blue reference corresponds to the radius of gyration of molecules in a corresponding completely quantum reference simulation. The shaded area marks the region used for the calculation of the RDF. Bottom: RDFs of the quantum-to-classical simulation calculated in the inner part of the quantum region (red), of a corresponding completely quantum reference simulation (blue), and of a full-classical system of particles interacting {\it via} the Silvera-Goldman potential (green). \label{fig2}}
\end{figure}

A snapshot of the dual-resolution simulation is presented in Fig. \ref{fig1}: the gradual change in size of the ring polymers indicates the transition from the classical to quantum mechanical regions and {\it vice versa}. Results are reported in Fig. \ref{fig2}.
The radius of gyration of the ring polymers in the quantum region (QM) perfectly reproduces the one of a corresponding fully quantum simulation. In the CL region the radius of gyration drops by $\approx 90\%$, indicating the classical character 
of the molecules (see Fig. \ref{fig1}). By means of the FEC a nearly flat density profile was obtained in the quantum region.

A quantitative measure of the fluid structure is provided by the radial distribution function (RDF): the latter is computed only in the inner part of the QM region (shaded region in Fig. \ref{fig2}).
In spite of the remarkable differences between the quantum fluid and the classical model, the RDF measured in the QM region matches very well the one obtained in the completely quantum reference simulations. These results show that in the QM region the quantum-to-classical coupling scheme correctly reproduces the structure of the quantum mechanical system.

\subsection{Speedup over full-quantum simulations} \label{speedup}

As mentioned earlier, in the proposed quantum-to-classical coupling scheme, interactions in the classical region do not need to be calculated $P$ times, with $P$ being the Trotter number, but because of the collapse of the polymer rings they are computed only once between the centers of mass of the (quasi point-like) rings. Additionally, a numerically simpler potential with a shorter cutoff is used in the classical region. Therefore, simulations become computationally more efficient, as we demonstrate hereafter.

Since in practice the method is most beneficial for systems in which the classical region is much bigger than the quantum region, we will consider such a situation. We perform 4 sets of simulations for different box sizes with each set consisting of a full-quantum, a full-classical and adaptive simulation in which the quantum region has a width of $2.0\,\text{nm}$ and the adjacent hybrid regions each have widths of $1.0\,\text{nm}$. The total box sizes as well as the corresponding molecule numbers for the simulations are presented in Tab. \ref{fig_supp_3}. 
The temperature and density are the same as before. In all cases, the classical regions are significantly larger than the quantum ones. Note that, here, ``classical simulation'' denotes a simulation of a WCA-liquid of collapsed polymer rings, exactly as in the classical region of the adaptive simulations, and not of a classical, i.e. with $P=1$, liquid of parahydrogen. All simulations are run for 400 sweeps and in the case of the adaptive simulation, a FEC is applied. Furthermore, the set of Monte Carlo moves is chosen for all of them to be that previously used for the adaptive simulations.

\begin{table}[!ht]\renewcommand{\arraystretch}{1.3}\addtolength{\tabcolsep}{-1pt}\renewcommand{\tabcolsep}{0.3cm}
\begin{center}
\begin{tabular}{c c c c}
\toprule[0.03cm]
Number of molecules & $L_x$ & $L_y$ & $L_z$  \\ 
\midrule[0.03cm]
4964 & $24.0\,\text{nm}$ & $3.123\,\text{nm}$ & $3.123\,\text{nm}$ \\
6619 & $32.0\,\text{nm}$ & $3.123\,\text{nm}$ & $3.123\,\text{nm}$  \\
8273 & $40.0\,\text{nm}$ & $3.123\,\text{nm}$ & $3.123\,\text{nm}$  \\
9928 & $48.0\,\text{nm}$ & $3.123\,\text{nm}$ & $3.123\,\text{nm}$\\
\bottomrule[0.03cm]
\end{tabular}
\caption{Number of molecules and box geometries for the different sets of simulations for the calculation of the computational gain of adaptive quantum-classical over full-quantum simulations.}
\label{fig_supp_3}
\end{center}
\end{table}

\begin{figure}[h]
  \includegraphics[width=\columnwidth]{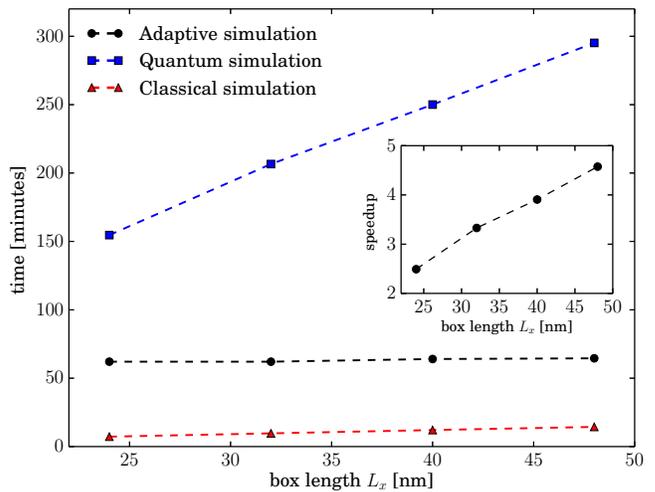}
  \caption{Main figure: times required for the potential energy calculations in quantum, classical and adaptive simulations for systems of four different box lengths $L_x$. Inset: speedup of the adaptive quantum-classical simulations. The lines are a guide to the eye. 
	\label{fig_supp_4}}
\end{figure}

In general, the overall speedup of the simulations strongly depends on the details of the implementation of the algorithm and is therefore platform dependent. For example, if there is a high overhead in the code, the overall computational gain by more efficient potential energy calculations will be small. If the program spends most of its time with these calculations, a significant improvement is possible. Hence, in order to obtain platform independent results, we only measure the time our code spends with potential energy calculations. These times are plotted in Fig. \ref{fig_supp_4}. Additionally, the corresponding speedups, defined as $T_\text{quantum}/T_\text{adaptive}$ with $T_\text{adaptive}$ ($T_\text{quantum}$) being the time spent for the energy calculations in the adaptive (quantum) simulations, are presented.

It can be seen that the adaptive simulations are significantly faster than their corresponding fully quantum counterparts. For the largest box, the energy calculations in the adaptive quantum-classical simulations are faster by a factor of $\approx4.5$ than the full-quantum simulations. Furthermore, it is visible that the time required for the adaptive simulations stays nearly constant for the different box sizes. The reason for it is that the computational cost of the interactions between classical molecules in the larger simulations is negligible compared to the time required for the computation of the potential energies in the quantum region.

\section{Conclusions} \label{conclusions}

In conclusion, we have derived a bottom-up Hamiltonian-based path integral formulation of a system of atoms or molecules whose quantum character depends on spatial location, and smoothly changes as the atoms or molecules diffuse.
The formalism is derived with the aim of providing a new approach for treating quantum condensed-phase soft-matter problems at multiple levels of resolution, here, employing both quantum and ``classical'' regions.  Possible future applications are diverse and include, for example, adaptive quantum-classical simulations of interface systems, membranes, and proteins. The approach will also allow rigorous treatment of the quantum grand-canonical ensemble. Due to the reduced number of degrees of freedom in the classical subdomain, the protocol presented enables a computationally more efficient sampling of configurations compared to a fully quantum simulation. This, in turn, allows an extension of the accessible time- and length-scales. Furthermore, the proposed scheme can also be employed in more advanced PI simulation techniques, such as Centroid Path Integral MD \cite{tuckermann_book,cmd1,cmd2} or Ring Polymer MD \cite{tuckermann_book,rpmd}. The development of such applications is the goal of a future study.

\section{Acknowledgments}

K. Kreis is recipient of a fellowship funded through the Excellence Initiative (DFG/GSC 266). Funding from the SFB--TRR 146 grant is gratefully acknowledged.

\bibliographystyle{rsc}
\bibliography{bibliography}

\end{document}